\documentclass[twocolumn,showpacs,preprintnumbers,prl,amsmath,amssymb,superscriptaddress ]{revtex4}

\usepackage{amsmath}    
\usepackage{amssymb}
\usepackage{graphicx}   
\usepackage{array}
\usepackage{comment}
\usepackage{epstopdf}

\newcommand{\etal}{{\it et al.}~}

\newcommand{\SrNi}{SrFe$_{2-x}$Ni$_x$As$_2$ }
\newcommand{\SrPt}{SrFe$_{2-x}$Pt$_x$As$_2$ }

\newcommand{\SrCo}{SrFe$_{2-x}$Co$_x$As$_2$ }
\newcommand{\BaCo}{BaFe$_{2-x}$Co$_x$As$_2$ }
\newcommand{\BaNi}{BaFe$_{2-x}$Ni$_x$As$_2$}
\newcommand{\BaRu}{BaFe$_{2-x}$Ru$_x$As$_2$ }

\newcommand{\SrFeAs}{SrFe$_2$As$_2$ }
\newcommand{\BaFeAs}{BaFe$_2$As$_2$ }
\newcommand{\CaFeAs}{CaFe$_2$As$_2$ }
\newcommand{\ie}{{\it i.e.}}

\newcommand{\Tc}{$T_c$~}
\newcommand{\Tcmax}{$T_{c(max)}$~}
\newcommand{\Tcz}{$T_{c0}$~}
\newcommand{\xten}[1]{$\times 10^{#1}$}

\begin{document}

\title{Universal pair-breaking in transition metal-substituted iron-pnictide superconductors }

\author{Kevin Kirshenbaum, S. R. Saha, S. Ziemak, T. Drye, J. Paglione}
 \email{paglione@umd.edu}
 \affiliation{Center for Nanophysics and Advanced Materials, Department of Physics, University of Maryland, College Park, MD 20742}

\date{\today}

\begin{abstract}

The experimental transport scattering rate was determined for a wide range of optimally doped transition metal-substituted FeAs-based compounds with the ThCr$_2$Si$_2$ (122) crystal structure. The maximum transition temperature \Tc for several Ba-, Sr-, and Ca-based 122 systems follows a universal rate of suppression with increasing scattering rate indicative of a common pair-breaking mechanism. Extraction of standard pair-breaking parameters puts a limit of $\sim$26~K on the maximum \Tc for all transition metal-substituted 122 systems, in agreement with experimental observations, and sets a critical scattering rate of $1.5\times 10^{14}$~s$^{-1}$ for the suppression of the superconducting phase. The observed critical scattering rate is much weaker than that expected for a sign-changing order parameter with strong interband scattering, providing important constraints on the nature of the superconducting gap in the 122 family of iron-based superconductors.

\end{abstract}


\maketitle


The discovery of iron-based superconductors in 2008 breathed new life into the study of high temperature superconductivity, with numerous families of compounds since discovered, characterized and extensively studied \cite{Reviews}. Intermetallic iron-based systems with the ThCr$_2$Si$_2$ ``122'' structure and doped with transition metal (TM) substitution on the iron site remain the most widely studied due to the feasibility of synthesizing large single-phase crystals coupled with the ability to substitute a plethora of TM elements for iron. With superconductivity induced by substuting almost any of the TM elements in the Fe, Co, and Ni columns, the robustness of these superconductors to disorder -- in particular, disorder focused directly in the active pairing layer -- provides a striking constrast to the sensitivity found in other unconventional superconductors. Furthermore, this robustness initiated one of the early challenges to the proposed $s_\pm$ sign-changing gap symmetry \cite{MazinKuroki}, and has been toted as evidence for a non-sign-changing $s$-wave pairing symmetry \cite{Onari}.

The role of TM substitution in both promoting a superconducting state and shaping the phase diagrams of the 122 systems is an important topic of ongoing debate. 
In the Ba-based 122 systems, the substitution-induced positioning of the superconducting phase scale reasonably well with d-electron count (with the exception of Cu substitution) \cite{CanfieldAnnRev}, and ample evidence of modifications to band structure, carrier concentrations and magnetic interactions support a rigid band shift doping model \cite{Reviews}. However, theoretical models predicting the localization of added $d$-electrons and the importance of impurities raise questions about this approach \cite{Wadati157004,Nakamura144512}. Moreover, the similarity of the phase diagram produced by nominally isovalent Ru substitution \cite{Schnelle214516, Sharma174512} to that of its aliovalent counterparts necessitates a better understanding of the true nature of TM substitution.

In this study we compare the elastic transport scattering rate and superconducting transition temperature \Tc for a wide range of optimally doped TM-substituted 122 compounds and observe a universal correlation that follows an Abrikosov-Gor'kov (AG)-like pair-breaking suppression for all types of transition metal substituents and alkaline earth cations.
We show that the large variations in optimal \Tc values found in different 122 systems are due to variations in impurity scattering rate, but are also limited by an ideal zero-scattering limit that lies much below the \Tc values of alkali metal-doped 122 systems. 
We discuss implications for order parameter structure as well as constraints on the inter- and intraband coupling strength determined by the universal relation.

Single crystals of AFe$_{2-x}$TM$_x$As$_2$ compounds (with A=Ba, Sr; TM=Co, Ni, Pd, Pt) were synthesized using the FeAs self-flux method described previously \cite{Saha224519}.  
TM concentrations were determined by wavelength dispersive x-ray spectroscopy (WDS). 
Resistivity and Hall effect data were measured in a commercial cryostat with Hall coefficient ($R_H$) values obtained by antisymmetrizing field sweeps at constant temperature between -5~T and +5~T. To minimize geometric factor errors in determining scattering rates, six-wire measurements were used to simultaneously determine longidudinal ($\rho_{xx}$) and transverse ($\rho_{xy}$) resistivities, using gold wire and silver paint contacts with typical contact resistances of $\sim 1~\Omega$.  Transition temperatures for samples measured by our group were determined by the mid-point of the resistive transitions, and are well documented in previous publications to coincide with magnetization measurements \cite{Saha224519, Kirshenbaum144518}. Data obtained from the literature utilized the same criteria when possible, and used stated values otherwise.

\begin{figure}[!]
    \includegraphics[width = 3.5in]{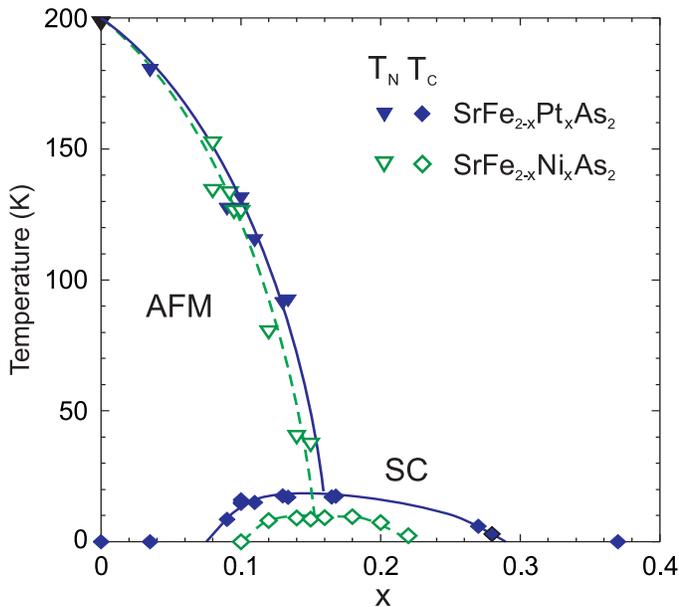}
    \caption{\label{phasediag} Phase diagrams of the \SrNi \cite{Saha224519} and \SrPt \cite{Kirshenbaum144518} systems. Antiferromagnetic (triangles) and superconducting (diamonds) transition temperatures are plotted for Ni- (open symbols) and Pt-doped (closed symbols) systems.  The similar rate of suppression of the magnetic phases and the position of the superconducting domes, with optimal doping at x $\simeq 0.16$ for both cases, is to be contrasted with the considerably different \Tcmax values of 9~K and 16~K for Ni- and Pt-doped series, respectively.}
\end{figure}


Unlike the \BaFeAs family, the maximum or ``optimal'' transition temperature \Tcmax for different TM-doped versions of the \SrFeAs system exhibits a wide variation of values, reaching $\sim$20~K for Co, Rh and Ir ~\cite{LeitheJasper207004, Han024506}, 16~K for Pt \cite{Kirshenbaum144518}, 9~K for Ni \cite{Saha224519} and 8~K for Pd substitution \cite{Han024506}. 
As a prime example, we directly compare the phase diagrams of the \SrNi \cite{Saha224519} and \SrPt \cite{Kirshenbaum144518} systems in Fig.~\ref{phasediag}. As shown, the antiferromagnetic order transition $T_N$ follows an almost identical decline as a function of either Pt or Ni substitution, with minimal difference between the two systems. The similar positioning of the superconducting dome for each system at an optimal concentration of $x \simeq 0.16$ follows that expected for the nominally equivalent addition of two $d$-electrons from both Pt and Ni substituents, as compared to that of \SrCo with only one $d$-electron contribution and a significantly larger optimal doping of $x \simeq 0.24$ \cite{Hu094520}.
However, a significant factor of two difference is apparent in \Tcmax values, presenting an intriguing contrast in two systems with nominally identical phase diagrams. 

With similar modification of unit cell parameters \cite{Kirshenbaum144518}, identical oxidation states and nearly identical phase diagrams in both substitution series, we consider intrinsic variations in pair-breaking scattering rates as the primary origin of this contrast.
Following previous studies, which have shown that electron bands dominate transport in the TM-doped systems \cite{Olariu054518, Fang140508, Kemper184516} and optical conductivity studies which indicate a single dominant Drude-like component \cite{Nakajima104528, Barisic054518}, we utilize a simple one-band model \cite{RA057001, Karkin174512} to estimate the elastic ($T=0$) transport scattering rate, allowing for broad comparisons between different systems and data sets.
The normal state scattering rate, 
$\Gamma = \frac{e \rho_{xx}}{m^* R_H}$, 
where $e$ is the electronic charge and $m^*$ is the effective mass, is determined from transport measurements by extrapolating the resistivity $\rho_{xx}$ and Hall coefficient $R_H$ to zero temperature using power law fits over an extended range of temperatures above $T_c$. We use an effective mass of $m^* = 2 m_e$ based on the measured values for the electron bands in CaFe$_2$As$_2$ \cite{Harrison322202}, SrFe$_2$As$_2$ \cite{Sebastian422203} and BaFe$_2$As$_2$ \cite{Terashima176402} from quantum oscillation measurements and an optical conductivity measurement on optimally Co-doped \BaFeAs\cite{Tu174509}.

\begin{figure}[!]
    \includegraphics[width = 3.25in]{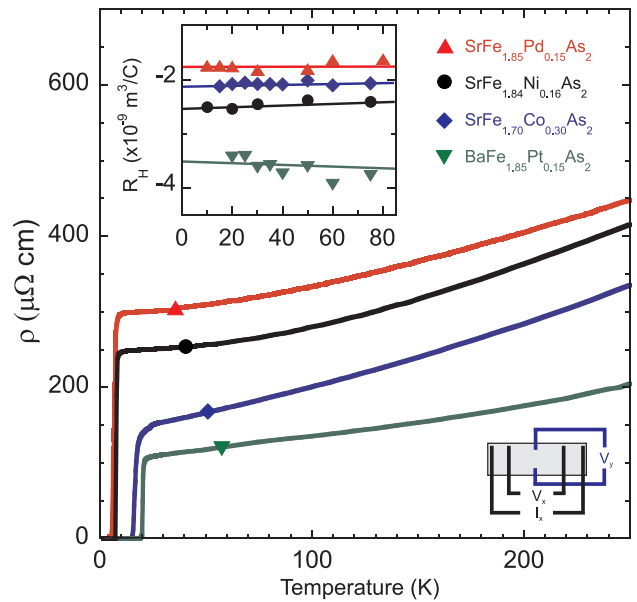}
    \caption{\label{rho-RH} Six-wire measurements of resistivity (main panel) and Hall coefficient (inset) of optimally-doped SrFe$_{2-x}$Pd$_x$As$_2$, SrFe$_{2-x}$Ni$_x$As$_2$, SrFe$_{2-x}$Co$_x$As$_2$ and BaFe$_{2-x}$Pt$_x$As$_2$, with \Tc values of 7, 8, 17 and 20~K, respectively.  The schematic depicts the configuration of the six-wire measurement.}
\end{figure}

Six-wire measurements were used to determine both $\rho_{xx}$ and $R_H$ simultaneously for several optimally-doped samples with a range of \Tc values, thereby significantly reducing geometric factor error in calculating $\Gamma$ by eliminating sample thickness dependence.
Shown in Fig.~\ref{rho-RH}, four different optimally-doped TM-substituted samples (Ni-, Pd-, Co- and Pt-based with \Tc values of 7~K, 8~K, 17~K, and 20~K, respectively) exhibit an observable difference in absolute resistivity values dominated by a rigid shift in the zero-temperature elastic contribution $\rho_0$, as evident from the comparable inelastic contributions (\ie, slope of $\rho(T)$). The resulting contrast in \Tc values follows this trend, with a systematic reduction of \Tc with increasing $\rho_0$. 

\begin{figure}[!]
    \includegraphics[width = 3.25in]{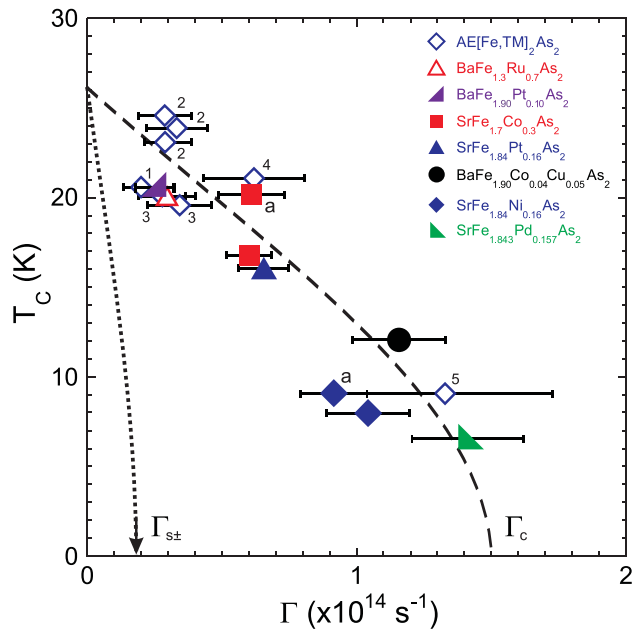}
    \caption{\label{pairbreak} Effect of transition metal substitution on \Tc values of a wide variety of 122 superconductors at optimal doping concentrations, plotted as a function of the experimental transport scattering rate $\Gamma \equiv e\rho/R_H m^{*}$ (see text). 
    Closed symbols indicate six-wire measurements (see text), and open symbols indicate values obtained from literature data for (1) CaFe$_{1.92}$Co$_{0.08}$As$_2$ \cite{Matusiak020510}, (2) \BaCo \cite{Mun09061548,RA057001}, (3) \BaNi \cite{Xu101016,Li025008}, (4) SrFe$_{1.74}$Co$_{0.26}$As$_2$ \cite{Zhang062510} and (5) SrFe$_{1.84}$Ni$_{0.16}$As$_2$ ($\rho_0$=212 $\mu\Omega$cm, $R_H$(0~K)=$1.4\times 10^{-9}$ m$^3$/C) \cite{Butch024518}. Samples of \SrCo and \SrNi denoted with an ``a'' are annealed (see text and footnote \cite{SrCo_footnote}).    
    All \Tc values follow a universal rate of suppression with $\Gamma$ well described by 
    an Abrikosov-Gor'kov fit (dashed line) that is much weaker than expected for a superconductor with $s_\pm$ symmetry and interband scattering (dotted line) \cite{Kogan214532, Karkin174512, Tarantini087002}.}
\end{figure}

We compare the resultant $T_c(\Gamma)$ values with those calculated for all optimally doped TM-substituted 122 samples with $\rho_{xx}$ and $R_H$ values available in the literature, as shown in Fig.~\ref{pairbreak}. (All data corresponds to systems with electron-dominated transport with the exception of Ru substitution, for which we utilize the electron component of $\rho_{xx}$ extracted with a two-band analysis and n$_e$ from ARPES measurements\cite{Rullier224503} to obtain a value of $\Gamma$ that can be compared with the other data.)
Remarkably, all \Tc values follow the same trend of suppression with increasing $\Gamma$, as expected in the AG formalism for a superconductor with increasing levels of pair-breaking impurities \cite{Abrikosov1243,Karkin174512,Kawamata11051000,Bang054529}, but surprising in light of the variety of systems presented. In particular, there is no clear trend associated with species of alkaline earth cation or transition metal substituent except for an average reduced scattering rate for Ba-based systems. This is likely correlated with both the lower substitution concentrations required to reach optimal doping as well as the lower $T_N$ ordering temperatures in \BaFeAs as compared to both \SrFeAs and \CaFeAs.  Note that $\Gamma$ values for \BaCo and \BaNi are nearly identical to those obtained in optical conductivity measurements \cite{Barisic054518} if we assume 
the same effective mass values, providing a confirmation of our analysis.

The rate of suppression of $T_c$, defined by the critical scattering rate $\Gamma_c$ where \Tc is completely suppressed, is in general dependent on the type of scatterers and the order parameter symmetry:  
according to Anderson's theorem, fully-gapped $s$-wave superconductors only respond to magnetic impurities, while unconventional pairing symmetries can be affected by both magnetic and nonmagnetic impurities \cite{Paglione703}. Assuming predominant non-magnetic scattering as evidenced by a paramagnetic normal state and no obvious indication of enhanced magnetism due to TM substitution ({\it e.g.}, absence of any enhanced susceptibility) \cite{Saha224519, Kirshenbaum144518},
the presence of non-magnetic pair-breaking points to a sign-changing order parameter.
However, several substitution \cite{Kawamata11051000, Li020513} and irradiation \cite{Nakajima10092848, Tarantini087002, Karkin174512} studies report a much weaker rate of suppression than that expected for a sign-changing order parameter; calculations for an ideal $s_\pm$ superconductor with full gaps on both bands \cite{Bang054529} and strong interband scattering yield $\Gamma_c(s_\pm)$ = 1.8\xten{13}~s$^{-1}$ \cite{Kogan214532, Karkin174512, Tarantini087002}, with similar values for the $d$-wave case \cite{Bang054529}.
Shown in Fig.~\ref{pairbreak}, a fit to the typical AG functional form \cite{Abrikosov1243} yields a value $\Gamma_c = 1.5\times 10^{14}$~s$^{-1}$ corresponding to a critical mean free path of $\sim 1.1$~nm (using Fermi velocity $v_f = 1.7 \times 10^5$~m/s \cite{Sebastian422203}), close to the expected superconducting coherence length $\xi=2.8$~nm \cite{Yin097002}. 

However, this value is also an order of magnitude weaker than the expected $\Gamma_c(s_\pm)$, presenting a significant challenge to models considering a fully gapped $s_\pm$ pairing symmetry, particularly in the presence of strong interband scattering \cite{Onari}. But calculations using the T-matrix approximation for an $s_\pm$ state emphasize that both inter- and intra-band scattering in the unitary limit can be decreased with appropriate parameters, resulting in a possible four-fold increase of $\Gamma_c$ \cite{Efremov11043840} that may offer an explanation, and may in fact be used to extract the relative strength of inter- and intraband scattering in these systems. In addition, recent studies that consider the effects of disorder on both superconductivity and competing states \cite{Fernandes} provide an alternative explanation for the apparent weak pair-breaking effects observed throughout the iron-based superconductor family.

The optimum clean-limit  ($\Gamma$=0) transition temperature \Tcz is an important parameter since it is the value that should be utilized in considering the intrinsic pairing strength. 
Our determination of \Tcz = 26~K is consistent with the well-established maximum \Tc value of $\sim 25$~K found among all TM-doped 122 systems \cite{Reviews}, as well as with extrapolated estimations of pressure- and doping-optimized systems such as shown in the comparison of \BaRu substitution and pressure dependence \cite{Kim11076034}. 
But this observation raises an intriguing question about why \Tcz does not approach that found in higher \Tc intermetallic systems including 
Sr$_{1-x}$K$_x$Fe$_2$As$_2$, Ba$_{1-x}$K$_x$Fe$_2$As$_2$ and BaFe$_2$As$_{2-x}$P$_x$
\cite{Reviews}, which have calculated $\Gamma$ values in the range shown in Fig.~\ref{pairbreak}. In contrast to the typical explanation of a reduced level of active-plane disorder as the reason for higher \Tc values in the alkali metal-doped systems, the determination of \Tcz and its failure to reach $\sim 40$~K suggests a fundamental asymmetry in pairing strength between electron- and hole-doped systems that does not arise from scattering differences alone (although the effects of strong scattering in the hole bands \cite{RA057001,Rullier224503,Olariu054518,Fang140508} cannot be discounted as a factor in the observed asymmetry).


A universal $T_c(\Gamma)$ relation suggests a similar pairing potential for all TM-doped 122 compounds that is disrupted by a common scattering mechanism.
It is not clear why certain TM substitutions induce more scattering than others, but dramatic variations in seemingly similar elemental substitutions are not unprecedented. For instance, the \BaRu system requires $\sim 30$-40\% Ru substitution to obtain optimal doping, which is almost four times higher concentration than Co substitution but results in a very similar value of \Tcmax. Such a contrast has been argued to arise from the aliovalent versus isovalent nature of, respectively, Co and Ru substituents, but recent work has put this into question. Mossbauer studies of \BaCo and \BaNi\ find no change in $d$-electron population with substitution \cite{Khasanov202201}, while an x-ray absorption study reveals no change in the Fe valence with Co substitution in \BaCo \cite{Bittar267402}. Furthermore, recent calculations suggest that substituted $d$-electrons can remain localized at the substituent sites \cite{Wadati157004}, either still resulting in a rigid band shift \cite{Nakamura144512} or generating a phase diagram strikingly similar to that expected from a rigid band shift \cite{Vavilov1110.0972}.

Variations in impurity or disorder levels due to details of substitution chemistry likely play a key role in explaining the variation in $\Gamma$ values observed in the 122 series of superconductors. This is corroborated by observations of enhancements in \Tc values after annealing crystals of both low- and high-$\Gamma$ systems, in particular \BaCo \cite{Gofryk} and \SrNi \cite{Saha_anneal}, respectively, and confirmed by our study of a \SrNi crystal with six-wire measurements obtained before and after annealing: as shown in Fig.~\ref{pairbreak}, the shift of data along the AG curve indicates an inverse relation between \Tc and $\Gamma$ \cite{SrCo_footnote}.
In the case of \BaCo \cite{Gofryk}, annealing was shown to enhance \Tc to a maximum value of 25~K, consistent with our determined \Tcz value. The reason why the \BaCo system is closest to the clean limit is not known, however a lack of observable disorder in Fe-As bond lengths in \BaCo may have provided an important insight \cite{Granado184508}; it would be interesting to perform the same study on high-$\Gamma$ systems to confirm this scenario.


In conclusion, we have demonstrated the existence of a universal pair-breaking relation for a wide range of optimally transition metal-doped 122 systems, suggesting a common scattering mechanism and pairing potential across the series. The rate of suppression of \Tc and the contrast between the optimum (zero-scattering) \Tcz value of $\sim$26~K and the higher \Tc values achieved in non-transition metal substitution series provides important constraints on the pairing symmetry and mechanism in the intermetallic iron-based superconductors.

\begin{acknowledgments}

The authors acknowledge A. Chubukov, I.~I. Mazin, L~Taillefer and M.~A. Tanatar for useful discussions, and P.C. Canfield for providing samples of BaFe$_{1.90}$Co$_{0.04}$Cu$_{0.03}$As$_2$.
This work was supported by AFOSR-MURI [FA9550-09-1-0603] and NSF-CAREER [DMR-0952716].

\end{acknowledgments}


\end{document}